\documentclass[%
 reprint, 
showpacs,preprintnumbers,
 amsmath,amssymb,
 aps,
]{revtex4-1}
\usepackage{footnote}
\usepackage{graphicx}
\usepackage{dcolumn}
\usepackage{bm}
\usepackage{slashed}
\usepackage{mathrsfs}
\usepackage{enumitem}
\usepackage[mathlines]{lineno}
\newcommand{\be}{\begin{equation}}
\newcommand{\ee}{\end{equation}} 
\newcommand{\bes}{\begin{equation*}}
\newcommand{\ees}{\end{equation*}}
\newcommand{\bra}{\langle}
\newcommand{\ket}{\rangle}

\newcommand\myeq{\mathrel{\stackrel{\makebox[0pt]{\mbox{\normalfont\tiny $|eQ_i|\mathcal{F}\to \infty$}}}{=}}}

\newcommand\myqeq{\mathrel{\stackrel{\makebox[0pt]{\mbox{\normalfont\tiny $e \mathcal{F} \to \infty$}}}{\sim}}}

\newcommand\myleqq{\mathrel{\stackrel{\makebox[0pt]{\mbox{\normalfont\tiny $|eQ_i|\mathcal{F}\to \infty$}}}{\leq}}}

\newcommand{\norm}[1]{\left\lVert#1\right\rVert}

\begin{document}

\preprint{APS/123-QED}

\title{Non-perturbative path integral quantization of the electroweak model:
             the Maxwell integration}

\author{M. P. Fry}
\affiliation{%
University of Dublin, Trinity College, Dublin 2, Ireland\\
}%




\date{\today}

\begin{abstract}
 The non-perturbative path integral quantization of the electroweak model is confronted with an apparent instability when integrating over the Maxwell potential $A_{\mu}$ due to the fast growth of the box graphs $AAAA$ and $AAAZ$ for large amplitude variations of $A_{\mu}$. $Z_{\mu}$ is from the vector part of the weak neutral current. These graphs are unavoidable because they are conditionally convergent and have to be isolated in the model's exact Euclidean one-loop effective action arising from its fermion determinants. A previous QED calculation of the large amplitude variation of its fermion determinant for a class of random potentials showed that the $AAAA$ box graph cancels in this limit. Using this result it is shown that within the electroweak model large amplitude variations of $A_{\mu}$ for fixed $Z_{\mu}$  in a superposition of these fields cancel the $AAAA$ and $AAAZ$ graphs, thereby removing an apparent obstacle to the model's non-perturbative quantization. A negative paramagnetic term in the remainder opposes the effective action's growth for such variations. Its calculation requires knowledge of the degeneracy of the bound states of a charged fermion in the four-dimensional magnetic fields generated by the functional measure of $A_{\mu}$.
\end{abstract}

\pacs{12.20.Ds, 12.15.-y, 11.10.Jj, 11.15.Tk}
\maketitle


\section{Introduction}\label{SECTION1}

The renormalizable electroweak model with its 24 adjustable parameters, including three massive Dirac neutrinos and their mixing, has so far accounted for a wealth of experimental data. Every aspect of the model should therefore be examined, including its non-perturbative sector. It is the aim of this paper to examine some aspects of this neglected sector. 

Non-perturbative information about any electroweak process resides in its representation as a functional integral over the fields contributing to the process. After spontaneous symmetry breaking this is a Euclidean path integral of the form
\begin{eqnarray}\nonumber
    \mathcal{I} &=& \mathcal{N}^{-1} \int d\mu(A) d\mu(Z) d\mu(W^{\pm}) d\mu(H) \prod_i [d\psi_i] [d\psi_i^{\dagger}] ~ \\\label{11}
    &&  \times  e^{\int d^4x \mathcal{L}(A,Z,W^{\pm},H,\psi, \psi^{\dagger})} \mathcal{F}(A,Z,W^{\pm},H,\psi, \psi^{\dagger}) \
\end{eqnarray}
\noindent
where $i=e, \mu, \tau,  \nu_e,  \nu_{\mu},  \nu_{\tau}, u, d, c, s, t, b$. Here $A, Z, W^{\pm}, H$ denote the Maxwell field, neutral and charged vector bosons and Higgs field, while the  $\psi_i$  denote the lepton and quark fields. The unitary gauge is chosen so that ghost fields are not required. $\mathcal{F}$ is a polynomial in the gauge and Higgs fields and the fermion fields specific to the process. $\mathcal{N}$ is a normalization constant defined in Section \ref{SUB3D}. The functional measures $d\mu$ are Gaussian in the indicated fields so that the electroweak model's Lagrangian $\mathcal{L}$ only contains interacting fields.

As $\mathcal{L}$ is quadratic in the quark and lepton fields they can be integrated out using the rules for integrating a Gaussian composed of Grassmann 4-component spinors \cite{29}. Neglecting mixing for the present this results in the following factorized determinants from the neutral and charged weak current for each quark family $i = u,d ; c,s; t,b$ \cite{1}

 \begin{widetext}
\begin{equation}  \label{12e}
\begin{split}
{\rm det} G^{-1}_{t_{3L}(i)= \frac{1}{2}} ~ {\rm det} G_{t_{3L}(i)= -\frac{1}{2}}^{-1} \times \det [1-\frac{g^2}{8}  G_{t_{3L}(i)=- \frac{1}{2}}^{} {\slashed W}^{-}&(1-\gamma_5)G_{t_{3L}(i)= \frac{1}{2}}  {\slashed W}^{+}(1-\gamma_5) ], 
\end{split}
\end{equation}
\end{widetext}

\noindent
where

\begin{equation} \label{13}
\begin{split}
{\rm det} G^{-1}_{t_{3L}(i)} =&  \\
 {\rm det} \Big[\slashed P + m_i - &eQ_i \slashed A - \frac{g}{2\cos \theta_W} \slashed Z (g_V^i - g_A^i\gamma_5)  + \frac{gm_i}{2M_W}H \Big].
\end{split}
\end{equation}

\noindent
Integration over the leptons gives the same result except that $W^+ \leftrightarrow W^-$ and $t_{3L}(i) \to -t_{3L}(i)$ in (\ref{12e}), where $i = \nu_e, e ; \nu_{\mu}, \mu; \nu_{\tau}, \tau$. $G_{t_{3L}(i)}$ is the propagator of fermion $i$ in the presence of the external potentials $A_\mu, Z_\mu$, and the Higgs field given by the inverse of the operator in brackets on the right-hand side of (\ref{13})\footnote{The determinants in Eqs.(4) and (5) in \cite{1} contain misprints: the factor $(8g^2)^{-1}$  should read $g^2/8$. In the last term in Eq.(6) $gm_i /(2M_W )$ should be replaced with $gm_i H/(2M_W )$.}. Here $m_i$ and $M_W$ are the fermion and $W$-boson masses; $e$ is the positron electric charge and $Q_i$ is the charge of fermion $i$ in units of $e$; $\theta_W$ is the weak angle and $g = e/\sin \theta_W$. The vector and axial-vector couplings are

\begin{eqnarray} \label{14}
g_V^i &= &t_{3L}(i) - 2Q_i \sin^2 \theta_W, \\ \label{15}
g_A^i &= & t_{3L}(i), \ 
\end{eqnarray}

\noindent
where $t_{3L}(i)$ is the weak isospin of fermion $i$. We have adopted the conventions and notations of \cite{2}. 

Quark mixing does not alter the determinants in (\ref{13}) that are the focus of this paper.
It does modify the last determinant in (\ref{12e}) contributed by the charged weak-vector current as reported in Section \ref{SUB3C}. Although mixing greatly complicates this determinant it does not modify the conclusions of this paper. Based on this result the three massive Dirac neutrinos' mixing is neglected here.

The determinants in (\ref{12e}) when written as $\Pi_i \exp[{ \ln \det}(i)]$ generate an effective action through the shift $\mathcal{L} \to \mathcal{L} + \sum_i{ \ln \det}(i)$ in the remainder of $\mathcal{L}$ in (\ref{11}) after the fermion integration. The sum over $i$ includes fermion generations and color degrees of freedom. 
Each determinant must be defined by factoring out its tadpole, self-energy, triangle and box graphs. These are assumed to be regularized, renormalized and made gauge invariant before inserting them in the above sum. Anomalies are assumed to be already cancelled in the sum over fermion generations. These steps are discussed in Sections \ref{SUB2A},\ref{SUB3A} and \ref{SUB3C}. Consequently there are no renormalization counterterms in the one-loop effective action as defined here. Going beyond one loop requires the introduction of ultraviolet regulators that are introduced in Section \ref{SUB2A} and remaining sections.

In \cite{3} it was asked whether the electroweak model can be non-perturbatively quantized and, in particular, whether any of the unexpanded functional integrals in (\ref{11}) over the gauge and Higgs fields converge\footnote{It may be asked whether the answers to these questions matter. Even deciding whether a power series expansion is asymptotic requires non-perturbative information. Recall that such a series places a precise bound on the remainder after terminating it, which is clearly a non-perturbative result. More broadly, it matters knowing whether an electroweak process that includes dynamical fermions can be non-perturbatively calculated.}. It was decided to approach these questions by integrating over the Maxwell field first after integrating over the fermions. This avoids immediate confrontation with the unmeasured shape of the Higgs potential. As the gauge field self-interactions in the interaction Lagrangian in (1) are quadratic in $A$ 
\cite{21} \footnote{ Since the calculation is non-perturbative the required $A$-dependent renormalization counterterms that should be present in the interaction Lagrangian are a priori unknown. These cannot be determined until the functional integral over A is computed, assuming it converges.} convergence depends on the large amplitude variations of the renormalized determinants with $A$. 

The process of defining the determinants introduces the box graphs $AAAA$ and $AAAZ$, where $Z$ is from the vector part of the weak neutral current as described in Section \ref{SECTION3}. These graphs confront the electroweak model with a potential instability when integrating over $A$. This is an example of the large field problem of a singular perturbation of a Gaussian functional measure \cite{4}, in this case $d\mu(A)$ in (\ref{11}). It is known that the $AAAA$ graphs cancel in the strong field limit of QED's Euclidean effective action for a class of random potentials \cite{3}. This is reviewed in Section \ref{SECTION2}. Based on this result it is shown in Section \ref{SUB3A} that the strong field limit of $A_{\mu}$ for fixed $Z_{\mu}$ in a superposition of these fields cancels the $AAAA$ and $AAAZ$ graphs. Other potentially destabilizing graphs are discussed in Section \ref{SUB3C}. A paramagnetic term in the one-loop effective actions of QED and the electroweak model opposing their growth for large amplitude variations of $A_{\mu}$ is discussed in Sections \ref{SUB2C} and \ref{SUB3B}.

Section \ref{SECTION4} summarizes our results. The Appendix completes a previous calculation of the strong-field dependence of the scalar QED determinant \cite{3} that is required in Section \ref{SECTION2}.


\section{Review of the strong field behavior of QED's effective action}\label{SECTION2}

\subsection{Preliminaries}\label{SUB2A}
Any one of QED's determinants contributed by a quark or charged lepton is obtained by setting $g=0$ in (\ref{12e}) and (\ref{13}) and subtracting ${ \ln \det}(\slashed P + m_i)$ in (\ref{13}) to give the formal expression ${ \ln \det}(1-eQ_i S \slashed A)$ normalized to $0$ at $e=0$. $S = (\slashed P + m_i)^{-1}$ is the free propagator for the fermion $i$. The process of defining this determinant begins by noting that the allowed potentials must support the gauge-fixed Gaussian measure $d\mu(A)$ in (\ref{11}) on $\mathcal{S}'(\mathbb{R}^4)$, the space of tempered distributions. These distributional, random potentials are smoothed by convoluting them with functions $f_{\Lambda}$ belonging to $\mathcal{S}(\mathbb{R}^4)$, the space of functions of rapid decrease:

\begin{equation} \label{21}
A_{\mu}^{\Lambda}(x) = \int d^4y ~ f_{\Lambda}(x-y)A_{\mu}(y).
\end{equation}

\noindent
Then $A_{\mu}^{\Lambda} \in C^{\infty}$ and hence is infinitely differentiable. As discussed in \cite{1,3} this smoothing process also introduces a gauge invariance preserving ultraviolet cutoff required to regulate QED. Thus, from the covariance of the measure $d\mu(A)$, $\int d\mu(A)~ A_{\mu}(x) A_{\nu}(y) = D_{\mu \nu}(x-y)$, where $D_{\mu \nu}(x-y)$ is the free photon propagator in a fixed gauge, obtain

\begin{equation} \label{22}
\int d\mu(A) ~ A_{\mu}^{\Lambda}(x) A_{\nu}^{\Lambda}(y) = D_{\mu \nu}^{\Lambda}(x-y).
\end{equation}

\noindent
The regularizing propagator $D_{\mu \nu}^{\Lambda}(x-y)$ has the Fourier transform $\hat D_{\mu \nu}(k) |\hat{f}_{\Lambda}(k)|^2$ with $\hat{f}_{\Lambda} \in C_0^{\infty}$, the space of $C^{\infty}$ functions with compact support such as $\hat{f}_{\Lambda}(k)=1$, $k^2 \leq \Lambda^2$ and $\hat{f}_{\Lambda}(k)=0$, $k^2 \geq n\Lambda^2$, $n >1$ \cite{1,3}. It should be clear that the random potentials $A_{\mu}$ are part of the functional measure perturbing $d\mu(A)$ and that they are measurable as (\ref{22}) illustrates. The $A_{\mu}^{\Lambda}$ will now replace $A_{\mu}$ everywhere in the functional integrals over $A_{\mu}$ except the measure $d\mu(A)$. In the following the superscript $\Lambda$ will be omitted with the understanding that $A_{\mu}$ is now a $C^{\infty}$ potential. Only when it encounters the measure does $\Lambda$ reappear.

The regularization and renormalization of ${\rm det}(1 - eQ_i S \slashed A)$ results in the renormalized determinant ${\rm det}_{\rm ren}$ \cite{5,6,7}, otherwise known as the Euclidean vacuum persistence amplitude,

\begin{equation}  \label{23}
{ \ln \det}_{\rm ren}(1 - eQ_i S \slashed A) = \frac{1}{2} \Pi_{AA} + \frac{1}{4}\Pi_{AAAA} + { \ln \det}_5(1-eQ_i S \slashed A),
\end{equation}

\noindent
where

\begin{equation}  \label{24}
{ \ln \det}_{5} = {\rm Tr} \left[ {\rm ln}(1-e Q_i S \slashed A) + \sum_{n=1}^4 \frac{(e Q_i S \slashed A)^n}{n} \right].
\end{equation}

The $\Pi_{AA}$ and $\Pi_{AAAA}$ terms contain the renormalized photon self-energy graph and the gauge invariant $\gamma \gamma$-scattering graphs, corresponding formally to ${\rm Tr}(eQ_i S \slashed A)^2$ and ${\rm Tr}(eQ_i S \slashed A)^4$, respectively. These are calculated from an expansion to $O(eQ_i)^4$ of the proper time representation of ${ \ln \det}(1-eQ_i\slashed A)$ that includes a second-order on-shell charge renormalization subtraction \cite{8}. More information on this is given by (\ref{39}) and (\ref{310}) below. This expansion also sets the tadpole and triangle graphs in $\Pi_A$ and $\Pi_{AAA}$ equal to zero as required by $C$-invariance. The four subtractions in brackets in (\ref{24}) remove all terms through $O(eQ_i)^4$ from ${\rm det}_5$. The gauge invariance of ${\rm det}_5$ requires that it depends only on $F_{\mu \nu}$.

The representation (\ref{24}) of ${ \ln \det}_5$  is defined only if the non-Hermitian operator $S\slashed A$ is a compact operator belonging to $\mathcal{I}_r$ , $r>4$. The trace ideal   $\mathcal{I}_r$ $(1 \leq r < \infty)$ is defined for those compact operators $T$ with ${\rm Tr}(T^{\dagger} T)^{r/2} < \infty$. This means that the eigenstates of $T$ are complete and square-integrable and that its complex eigenvalues are discrete, have finite multiplicity, and satisfy $\sum_n |\lambda_n|^r < \infty$. General properties of $\mathcal{I}_r$ spaces and the properties of determinants of operators belonging to these spaces may be found in  \cite{9,10,11,12}. By a theorem of Seiler and Simon \cite{5,6,7,9,13} $S\slashed A \in \mathcal{I}_r$, $r>4$ provided $m_i \neq 0$ and $A_{\mu} \in \cap_{r>4} L^r (\mathbb{R}^4)$, thereby validating (\ref{24}) for this class of potentials. This restriction means that $A_{\mu}(x)$ falls off at least as fast as $1/|x|$ for $|x| \to \infty$ \footnote{Because the large distance behavior of $A_{\mu}^{\Lambda}$ and $A_{\mu}$ are the same due to the choice of $f_{\Lambda}$ in (\ref{21}), the distributional, random connection $A_{\mu}$ must also fall off at least as fast as $1/|x|$. There is no evidence for or against this assumption to the author's knowledge as discussed in Section \ref{SUB3D}. If progress is to be made in understanding the asymptotic behavior of the one-loop effective action in QED and the electroweak model for large amplitude variations of $A_{\mu}$ using presently known mathematics then this assumption has to be made.}, that it has no poles or branch points for finite $x$ such as $|x-x_0|^{-\beta}$, $\beta>0$, and that $A_{\mu}(x)$ is finite at $x=0$. From here on we will denote an eigenvalue of $S\slashed A$ by $1/e_n$.

Since $S \slashed A \in \mathcal{I}_r$, $r>4$, $\sum_n (1/|e_n|)^{4+ \epsilon} < \infty$, $\epsilon > 0$, so that ${\rm det}_5$ is an entire function of $eQ_i$ \cite{9,10,11,12} of order $4$ \cite{14}. That is, ${\rm det}_5$ is analytic in $eQ_i$ in the entire complex $e$-plane with $|{\rm det}_5| < A \exp(K|eQ_i|^{4+\epsilon})$ for positive constants $A$ and $K$. Since ${\rm det}_5 =1$ for $eQ_i = 0$, ${\rm det}_5 >0$ for real values of $e$ since the zeros of ${\rm det}_5$ lie off the real $e$-axis when $m_i \neq 0$. Because ${\rm det}_5$ is an entire function of $eQ_i$ of order $4$ ${ \ln \det}_5$ can impact on the $\Pi_{AAAA}$ term in (\ref{23}) for large amplitude variations of $A_{\mu}$. We will return to this below.

\subsection{Means of calculation}\label{SUB2B}

To appreciate the full significance of (\ref{23}) more information on the strong field behavior of ${\rm det}_5$ is required. Such information is obtained from the following representation of ${\rm det}_{\rm ren}$ \cite{3} derived from Schwinger's proper time representation of $\det_{\rm ren}$ \cite{8}:

 \begin{widetext}
 \be \label{25}
 \begin{split}
{ \ln \det}_{\rm ren} &= 2 \int_0^{\infty} \frac{dt}{t} ~ \left[ {\rm Tr} (e^{-P^2t} - e^{-(P-eQ_i A)^2t} ) - \frac{e^2Q_i^2 ||F||^2}{192\pi^2} \right] e^{-tm_i^2} + \frac{1}{2} { \ln \det}_3 (1 + \Delta_A^{1/2} \frac{1}{2} e Q_i \sigma F \Delta_A^{1/2} ) \\
& \quad + (eQ_i)^2 \int_0^{\infty} dt ~  e^{-tm_i^2} \left[ \frac{1}{32\pi^2t} ||F||^2 - \frac{1}{2} {\rm Tr}( e^{-(P-eQ_i A)^2t} F_{\mu \nu} \Delta_A F^{\mu \nu} )  \right].
\end{split}
\ee
\end{widetext}

The first term in (\ref{25}) is twice the proper time definition of the scalar QED determinant with an on-shell charge renormalization subtraction, where $||F||^2 = \int d^4x ~ F_{\mu \nu} F^{\mu \nu}$.

In the second term $\Delta_A = [ (P - eQ_i A)^2 + m_i^2 ]^{-1}$ is the propagator of a charged scalar particle in the external potential $A_{\mu}$ and $\sigma_{\mu \nu} = [\gamma_\mu, \gamma_\nu]/(2i)$. The Euclidean $\gamma$-matrices are anti-Hermitian. The Hermitian operator

\begin{equation} \label{26}
T = \Delta_A^{1/2} \frac{1}{2} e Q_i \sigma F \Delta_A^{1/2},
\end{equation} 

\noindent
belongs to the trace ideal $\mathcal{I}_3$ if $F_{\mu \nu} \in \cap_{r>2} L^r(\mathbb{R}^4)$ \cite{3}. Therefore, its eigenstates are complete and square-integrable. Its eigenvalues $\{ \lambda_n \}_{n=1}^{\infty}$ are real, discrete with finite multiplicity, occur in pairs $\lambda_n, -\lambda_{n}$ and satisfy $\sum_{n=1}^\infty |\lambda_n|^3 < \infty$. Then the second term in (\ref{25}) can be expressed as 

\begin{equation} \label{27}
\begin{split}
{ \ln \det}_3(1+T) & = { \ln \det} [ (1+T)\exp(-T + \tfrac{1}{2}T^2) ] \\
			     & =  {\rm Tr}[ \ln (1+T) - T + \tfrac{1}{2} T^2 ] \\
			     & = \sum_{n=1}^{\infty} [ \ln(1-\lambda_n^2) + \lambda_n^2 ], 	
\end{split}
\end{equation}

\noindent
where the sum is over positive eigenvalues. Since ${ \ln \det}_3$ is real and finite, $\lambda_n <1$ for all $n$ and hence

\begin{equation} \label{28}
{ \ln \det}_3(1+ \Delta_A^{1/2} \tfrac{1}{2} eQ_i \sigma F \Delta_A^{1/2}) \leq 0,
\end{equation} 
\noindent
since $\ln (1-x^2) + x^2 \leq 0$ for $0 \leq x \leq 1$. The strong-field dependence of the eigenvalues will be examined in the next section.

The last term in (\ref{25}) is connected with charge renormalization and is positive due to QED's lack of asymptotic freedom. 
Specifically, the third term's contribution to the strong-field asymptotic behavior of $\ln \det_{\rm ren}$ is $|| eQ_iF||^2/(32 \pi^2) \times \ln (|eQ_i| \mathcal F/m_i^2)$, while the first term reduces this by a factor of $2/3$, resulting in an overall growth of $(\beta_1||eQ_iF||^2/4)\times \ln(|eQ_i| \mathcal F/m_i^2)$, where $\beta_1=1/(12 \pi^2)$ is the coefficient of the one-loop QED beta-function \cite{3}. This result coincides with the
 analysis in \cite{28} for the case of constant $F$.

Each term of the right-hand side of (\ref{25}) is separately gauge invariant and ultraviolet finite. The advantage of this representation is that ${ \ln \det}_{\rm ren}$ is divided into three tractable terms each of which can be estimated in the strong-field limit. In particular, the separation of the paramagnetic spin term, represented by $\det_3$, from the rest of the spinor QED determinant is achieved here. As will be seen in Section \ref{SECTION3}, (\ref{25}) is immediately extendable to include the weak neutral vector current.

\subsection{Results} \label{SUB2C}

Let $\mathcal{F}$ fix the amplitude of $F_{\mu \nu}$ in which case $\mathcal{F}$ has the dimension of $L^{-2}$. Since $eQ_i$ always multiplies $F_{\mu \nu}$ the natural strong-field scaling parameter is $|eQ_i|\mathcal{F}$. Then for the smoothed potentials introduced above and for each charged fermion \cite{3}

\begin{widetext}
\begin{equation} \label{29}
\begin{split}
{ \ln \det}_{\rm ren} \qquad   \myeq \qquad \left( \frac{1}{48\pi^2}(eQ_i)^2 ||F||^2  - \frac{N}{2}  \right)   \ln \left( \frac{ |eQ_i|  \mathcal{F}}{m_i^2}  \right) + R.
\end{split} 
\end{equation} 
\end{widetext}
$N$ in (\ref{29}) is contributed by the spin-dependent term ${\rm det}_3$ in $(\ref{25})$. It is the number of eigenstates of $T$ in (\ref{26}) having an eigenvalue $\lambda \nearrow 1$ as $|eQ_i|\mathcal{F} \to \infty$. In the absence of such eigenstates the remainder $R$ satisfies 

\begin{equation} \label{210}
\lim_{|eQ_i|\mathcal{F} \to \infty} \frac{R}{ (eQ_i \mathcal{F})^2 \ln( |eQ_i| \mathcal{F}) } = 0.
\end{equation}

\noindent
The result (\ref{29}) summarizes the results (6.44)-(6.46) in \cite{3}. We note that the inequality in those results has been replaced here with equality since the strong-field dependence of the scalar QED determinant required to obtain these results has been sharpened in the Appendix of this paper. The first term in (6.45) and (6.46) should be multiplied by $1/2$. The result (14) agrees with the asymptotic behavior of the 1-loop Heisenberg-Euler effective Lagrangian for the case of a constant magnetic field for which N=0 after introducing a volume cutoff. See \cite{extra2} and references therein.

The $N$-dependent term in (\ref{29}) is relevant to the functional integrability of QED. Inspection of (\ref{27}) shows that an eigenvalue $|\lambda| \nearrow 1$ as $|eQ_i|\mathcal{F} \to \infty$ will cause ${ \ln \det}_3$ to assume a large negative value that is enhanced if the degeneracy $N$ of the associated eigenstates is large. Understanding this stabilizing result will decide in Section \ref{SECTION3} whether it extends to the entire electroweak model. Hence, a review of its derivation in \cite{3} is warranted here.

The eigenvalue $\lambda$ and its associated eigenstates are obtained by transforming the eigenvalue equation $T | \lambda \ket = \lambda  | \lambda \ket$ into the equivalent equation

\begin{equation} \label{211}
\left[  (P-eQ_iA)^2 + \frac{eQ_i}{2\lambda} \sigma F \right] \psi_{\lambda,n} = -m_i^2 \psi_{\lambda,n},
\end{equation}

\noindent
where $\psi_{\lambda,n} \in L^2 (\mathbb{R}^4)$ and $n$ is the set of quantum numbers specifying the state. The state $\psi_{\lambda,n}$ will in general have both positive and negative chirality components. At this stage $\lambda$ is just one of a discrete set of eigenvalues $\{  \lambda_k\}_{k=1}^{\infty}$ of $T$ that result in a bound state with energy $-m_i^2$ for a fixed value $eQ_i$. Bound states are possible when $m_i \neq 0$, $0 < |\lambda| <1$, when $\lambda >0 (<0)$ and $eQ_i \langle \lambda, n | \sigma F | \lambda, n \rangle <  0 ( >0)$ due to the formation of sufficiently broad and deep potential wells. Assume $eQ_i > 0$, and that $0<\lambda<1$ following Section \ref{SUB2B}. Suppose the potential in (\ref{211}) also supports a zero mode $\psi_{0,n}$ that satisfies

\begin{equation} \label{212}
\left[  (P-eQ_iA)^2 + \frac{eQ_i}{2} \sigma F \right] \psi_{0,n} = 0.
\end{equation}

\noindent
The square-integrable state $\psi_{0,n}$ has definite chirality. Equation (\ref{212}) requires $\bra 0,n | \sigma F | 0,n \ket  < 0$. The state $| 0, n \rangle$ denotes a zero mode state with quantum numbers $n$ and not a state with $\lambda = 0$. Referring to (\ref{211}), $\bra \lambda,n | \sigma F | \lambda, n \ket < 0$. From (\ref{211}), (\ref{212}) there follows 

\begin{equation}\label{213}
\frac{\lambda}{1-\lambda} = \frac{ |eQ_i| \mathcal{F} }{2m_i^2} \left| \frac{ \bra 0,n | \sigma F | \lambda,n \ket }{\mathcal{F} \bra 0,n | \lambda,n \ket } \right|.
\end{equation}

\noindent
If all of the angular-momentum-like quantum numbers $n$ are the same and $|\lambda,n \ket$ has mixed chirality then $|0,n\ket$ projects out one of the chirality components from $|\lambda,n \ket$, and we expect $\bra 0,n | \lambda,n \ket \neq 0$. Based on our limited knowledge of four-dimensional Abelian zero-modes \cite{17} they have a distinctive structure, and so the non-vanishing of $\bra 0,n | \lambda,n \ket $ distinguishes $|\lambda,n \ket$ and its eigenvalue $\lambda$ from all the other eigenstates of $T$. A necessary condition on $F_{\mu \nu}$ to define $\det_3$ is $F_{\mu \nu} \in \cap_{r>2} L^r(\mathbb{R}^4)$ \cite{3}. Therefore, $F_{\mu \nu}$ is a bounded function and 

\begin{equation}\label{214}
\left|  \frac{ \bra 0,n | \sigma F | \lambda,n \ket }{\mathcal{F} \bra 0,n | \lambda,n \ket}  \right| \leq K,
\end{equation}

\noindent
where $K$ is bounded for large $\mathcal{F}$. Equation (\ref{213}) and (\ref{214}) imply $\lambda \nearrow 1$ as $|eQ_i|\mathcal{F} \to \infty$. The operator transforming $|\lambda, n\ket$ into the negative eigenvalue state $|-\lambda,n\ket$ is constructed in Appendix D of \cite{3}. Since the sum in (\ref{27}) is over $\lambda^2$ the negative eigenvalues are included in going from line 2 to line 3 in (\ref{27}). Insertion of (\ref{213}) and (\ref{214}) in (\ref{27}) then gives the $N$-dependent term in (\ref{29}). An analytic calculation of the eigenvalue $\lambda$ for a family of zero-mode supporting potentials is given in Section V and Appendix E of \cite{3}.

The foregoing leads to the general statement: If the potential $A_{\mu}$ also supports a zero mode state $| 0,n\ket$ and one of the positive eigenvalue states $|\lambda,n\ket$ of $T$ has the same quantum numbers $n$ as $|0,n\ket$ and $\bra 0, n|\lambda, n \ket \neq 0$, then $\lambda \nearrow 1$ as $|eQ_i|\mathcal F \to \infty$. An operator can be constructed that maps $|\lambda, n \ket$ to the orthogonal state $|-\lambda, n\ket$.

$N$ in this case is the number of states $|\lambda, n\ket$ and is also equal to the number of zero modes $|0,n\ket$  as these two sets of states are in one-to-one correspondence. This line of reasoning makes it clear that the mass singularity contributed by ${ \ln \det}_3$ to ${ \ln \det}_{\rm ren}$ cannot be removed. This is unlike the mass singularity associated with the first term in (\ref{29}) that can be removed by renormalizing off-shell.

If the zero mode supporting potential $A_{\mu}$ falls off as $1/|x|$ for $|x|\to \infty$ and all of the zero modes have the same chirality then their number, $N$, is given by the absolute value of the chiral anomaly, $(eQ_i)^2 | \int d^4x ~ \epsilon_{\mu \nu \alpha \beta} F^{\alpha \beta}F^{\mu \nu} | / (32\pi^2)$ \cite{15,16} and $R$ in (\ref{29}) satisfies (\ref{210})\footnote{ Strictly, $N$ is the greatest integer value of the absolute value of the chiral anomaly. The remainder is the contribution from the zero-energy scattering phase shifts. If the chiral anomaly is precisely an integer then $N$ in (\ref{29}) is replaced with $N-1$ \cite{15,16}.}. In this case $F_{\mu \nu}$ is not square-integrable, requiring a volume cut-off in $||F||$ in (\ref{29}) that is discussed in Section \ref{SUB3D}. The presence of $||F||$ in (\ref{29}) is from a charge renormalization subtraction \cite{3} and is independent of $\det_5$ which has no divergence for the class of fields under consideration here.

If the zero modes do not have the same chirality then the Atiyah-Singer index theorem generalized to non-compact Euclidean spacetime \cite{15,16} no longer gives their total number, and no bound can be placed on $R$. At present there is no evidence that a zero-mode supporting potential in four-dimensional QED can have zero modes with different chirality. In the single known case of such a potential with a $1/|x|$ falloff all of its zero modes are found to have the same chirality \cite{17}.

We have considered the states $|\lambda, n \ket$ obtained from (\ref{211}) that have eigenvalue $|\lambda| \nearrow 1$ as $|eQ_i|\mathcal{F} \to \infty$ when the potential also supports a zero mode. This was done because it allowed us to count the states $|\lambda,n\ket$  under the limitations discussed above. We see no reason why other admissible potentials cannot also produce eigenstates $|\lambda,n\ket$ from (\ref{211}) such that $|\lambda| \nearrow 1$ as $|eQ_i|\mathcal{F} \to \infty$. This opens the possibility of a much larger class of admissible potentials supporting $d\mu(A)$ that can result in an increasing $F_{\mu \nu}$-dependent degeneracy parameter $N$ in (\ref{29}). These potentials may be more likely to support $d\mu(A)$ than the highly restricted zero-mode supporting potentials, and, if so, will have a direct bearing on the convergence of the Maxwell integration in (\ref{11}) when $g=0$. This possibility was not noticed in \cite{3}. 

In fact, ${ \ln \det}_3$ in (\ref{25}) and (\ref{27}) may be the controlling term in ${ \ln \det}_{ren}$ for large variations of $F_{\mu \nu}$ for reasons discussed at the end of Section \ref{SUB3B} that are also applicable to QED.

\subsection{${\rm det}_5$ and $\Pi_{AAAA}$ } \label{SUB2D}

Assume that the zero modes, if any, supported by an admissible potential have the same chirality. Then the results (\ref{29}) and (\ref{210}) apply. Since the $\Pi_{AAAA}$ term in (\ref{23}) is of $O(eQ_i\mathcal{F})^4$ then $\det_5$ always cancels $\Pi_{AAAA}$ in the limit $|eQ_i|\mathcal{F} \to \infty$ to give the result in (\ref{29}).

Considering the complexity of $\Pi_{AAAA}$ when reduced to its gauge invariant form \cite{18,19} it is remarkable that the eigenvalues $\{ 1/e_n \}_{n=1}^{\infty}$ of $S \slashed A$ arrange themselves in $\det_5$ to cancel it in the strong-field limit, especially since

\begin{equation}\label{215}
\Pi_{AAAA} \neq -\frac{1}{4} \sum_n(1/e_n)^4.
\end{equation}

\noindent
To cancel $\Pi_{AAAA}$ and satisfy (\ref{29}) $\det_5$ must assume its allowed exponential growth, $A\exp(K|eQ_i|^4 \mathcal F^4)$, on the real $e$-axis.

We have no information on the relative sign of $\Pi_{AAAA}$ and ${ \ln \det}_5$ for a particular background field. In the preceding paragraph it is assumed that $\Pi_{AAAA}<0$. If $\Pi_{AAAA}>0$ then ${ \ln \det}_5$ must vary as $-(eQ_i\mathcal{F})^4$ on the real $e$-axis for large field fluctuations. Our analysis cannot distinguish between these cases, but it does rule out $\Pi_{AAAA}$ and ${ \ln \det}_5$ having the same sign when $|eQ_i|\mathcal{F}\to \infty$. In \cite{17} it was found that the large mass expansion of $\Pi_{AAAA}$ can change sign with different fields $F_{\mu \nu}$.

To go a step further and declare $\det_5$ an entire function of order $4$ and finite type would require that $\ln|\det_5|$ grows no faster than $  |e|^4(Q_i \mathcal F)^4$ along all rays in the complex $e$-plane. Ruling out growth such as $ |e|^4(Q_i \mathcal F)^4 \times \ln^\alpha(|eQ_i| \mathcal F)$, $\alpha>0$, along some rays requires sufficient symmetry in the distribution of the eigenvalues of $S\slashed A$ \cite{14}. Euclidean $C$-invariance\footnote{In this limited context we mean there exists a matrix $C$ such that $C\gamma_\mu C^{-1} = \gamma_{\mu}^T $. In the representation of the $\gamma$-matrices used in \cite{3}, Eq. (D7), $C = \gamma_3 \gamma_1$.} and the reality of $\det_5$ for real $e$ require these to occur in quartets $\pm e_n, \pm \bar e_n$ or as complex conjugate pairs. This may or may not be sufficient for $\det_5$ to be of finite type. 

Since $\det_5$ is an entire function of $eQ_i$ of order $4$ then by ($\ref{23}$) so is $\det_{\rm ren}$. Result (\ref{29}) shows that $\det_{\rm ren}$ - the Euclidean vacuum persistence amplitude - does not assume its maximal growth on the real $e$-axis. This confirms a long-standing conjecture of Balian, Itzykson, Parisi and Zuber \cite{20}.


\section{Extension of Section \ref{SECTION2}'s results to the electroweak model }\label{SECTION3}

\subsection{Cancellation of $\Pi_{AAAA}$ and $\Pi_{AAAZ}$} \label{SUB3A}

The relevance of the preceding results to the electroweak model becomes evident on referring to the determinants in (\ref{12e}) and (\ref{13}) and noting the superposition

\begin{equation*}
eQ_i A_{\mu} + \frac{g_V^ig}{2\cos \theta_W}Z_{\mu}.
\end{equation*}

\noindent
This suggests that the potential $V_{\mu}$ defined by

\begin{equation}\label{31}
eQ_iV_{\mu} = eQ_i \left( A_{\mu} + \frac{g_V^i}{2Q_i \cos \theta_W \sin \theta_W} Z_{\mu}  \right),
\end{equation}

\noindent
will be useful to study the interference of $A_{\mu}$ with $Z_{\mu}$ and the cancellation of the potentially destabilizing box graph $AAAZ$ for large amplitude variations of $A_{\mu}$. The relation $g=e/\sin \theta_W$ was used in (\ref{31}). Consider the formal operations on (\ref{13}):

 \begin{widetext}
 \be \label{32}
 \begin{split}
\ln \det \left( \slashed P - eQ_i \slashed V + m_i + \frac{g_A^ig}{2\cos \theta_W} \slashed Z \gamma_5 + \frac{gm_i}{2M_W}H \right) - \ln \det (\slashed P - eQ_i \slashed V + m_i) + \ln \det (\slashed P - eQ_i  &\slashed V + m_i) - \ln \det (\slashed P + m_i) \\
 = \ln \det (1-eQ_iS\slashed V) + \ln \det \left( 1 + S_V \left( \frac{g_A^ig}{2\cos \theta_W} \slashed Z \gamma_5 + \frac{gm_i}{2M_W} H  \right) \right),&
\end{split}
\ee
\end{widetext}

\noindent
where $\ln(\slashed P + m_i)$ is subtracted so that the right-hand side of (\ref{32}) vanishes when $e,g=0$. $S_V$ is the propagator of a charged fermion in the external potential $V$:

\begin{equation} \label{33}
S_V = (\slashed P - eQ_i\slashed V + m_i)^{-1}.
\end{equation}

\noindent
We will return to the last determinant in (\ref{32}) in Section \ref{SUB3C} below \footnote{The determinant decomposition in (3) in \cite{1} is now superseded by that in (\ref{32}). This has no effect on the results in \cite{1}.}. 

Our interest here is $\ln \det(1-eQ_iS\slashed V)$. It can be connected to the results for $\ln \det_{\rm ren}(1-eQ_iS\slashed A)$ with the shift $eQ_iA_{\mu} \to eQ_iV_{\mu}$ following (\ref{31}). It is assumed that $Z_{\mu}$ has been smoothed and made $C^{\infty}$ by the same procedure as in Section \ref{SUB2A} and that $Z_{\mu} \in \cap_{r>4} L^{r}(\mathbb{R}^4)$ as does $A_{\mu}$. The smoothing function $f_{\tilde \Lambda}$ for $Z$ should have $\tilde \Lambda \neq \Lambda$ to keep the regularizations relating to $A$ and $Z$ separate. The Seiler-Simon theorem in Section \ref{SUB2B} now applies to $S\slashed V$ so that this operator belongs to $\mathcal{I}_r$, $r>4$. Then representation (\ref{25}) for $\ln \det_{\rm ren}(1-eQ_i\slashed A)$ extends to the electroweak model on replacing $eQ_i A_{\mu}$ with $eQ_i V_{\mu}$:

 \begin{widetext}
 \be \label{34}
 \begin{split}
{\ln \det}_{\rm ren} (1-eQ_iS \slashed V) & = 2 \int_0^{\infty} \frac{dt}{t} ~ \left[ {\rm Tr}\left(e^{P^2t} - e^{-(P-eQ_iV)^2t}\right) - \frac{1}{192\pi^2} \norm{eQ_i F_{\mu \nu} + \frac{g_V^ig}{2\cos \theta_W} Z_{\mu \nu} }^2  \right]  e^{-tm_i^2} \\
							& \quad + \frac{1}{2} \ln {\det}_3\left( 1 + \Delta_V^{1/2} \frac{1}{2} \sigma_{\mu \nu} \left( eQ_i F^{\mu \nu} + \frac{g_V^ig}{2\cos \theta_W} Z^{\mu \nu}  \right) \Delta_V^{1/2}  \right) \\
							& \quad + \int_0^{\infty} dt ~ e^{-tm_i^2} \Bigg[  \frac{1}{32\pi^2t} \norm{ eQ_i F_{\mu \nu} + \frac{g_V^ig}{2\cos \theta_W} Z_{\mu \nu}   }^2 \\
							& \quad - \frac{1}{2}{\rm Tr} \left(  e^{-(P-eQ_iV)^{2}t} \left(eQ_i F_{\mu \nu} + \frac{g_V^ig}{2\cos \theta_W}Z_{\mu \nu} \right) \Delta_V \left( eQ_iF^{\mu \nu} + \frac{g_V^ig}{2\cos \theta_W} Z^{\mu \nu} \right)    \right) \Bigg].
\end{split}
\ee
\end{widetext}

\noindent
The propagator $\Delta_A$ has been replaced with the scalar propagator in the background potentials $A_{\mu}, Z_{\mu}$:

\begin{equation} \label{35}
\Delta_V = [(P-eQ_iV)^2 + m_i^2]^{-1}.
\end{equation}

The first term in (\ref{34}) is the scalar QED determinant in (\ref{25}) shifted to give the renormalized one-loop effective action of a charged particle propagating in the neutral vector potential $V_{\mu}$. The trace term is positive by Kato's inequality \cite{24,25,26,27}, which means that on average the energy levels of a scalar particle minimally coupled to a neutral vector potential increase. The remaining renormalization subtraction causes the first term to turn negative for $|eQ_i|{\mathcal{F}} \to \infty$. When combined with the leading positive renormalization subtraction in the third term in (\ref{34}) the result is a fast growing contribution to $\ln \det_{\rm ren}$ as seen in (\ref{312}) below.

In the second term the Hermitian operator

\begin{equation} \label{36}
T_V = \Delta_V^{1/2} \frac{1}{2} \sigma_{\mu \nu} \left( eQ_iF^{\mu \nu} + \frac{g_V^ig}{2\cos \theta_W}Z^{\mu \nu} \right) \Delta_V^{1/2}
\end{equation}

\noindent
belongs to the trace ideal $\mathcal{I}_3$ if $F_{\mu \nu}, Z_{\mu \nu} \in \cap_{r>2} L^r (\mathbb{R}^4)$ following a straightforward generalization of the result in Appendix A of \cite{3} by replacing $A_{\mu}$ with $V_{\mu}$ and $F_{\mu \nu}$ with $V_{\mu \nu}$. All of the properties of $\ln \det_3(1+T)$ and the eigenvalues of $T$ in (\ref{27}) carry over unchanged when $T$ is replaced with $T_V$. Therefore, the second term in (\ref{34}) is negative and can significantly reduce the growth of $\ln \det_{\rm ren}$ if one of the eigenvalues of $T_V$ approaches unity for large amplitude variations of $A_{\mu}$. This will be dealt with in Section \ref{SUB2B} below. 

For our purpose here we also introduce the alternative representation of $\ln \det_{\rm ren}$ following (\ref{23}), (\ref{24}):

\begin{equation} \label{37}
\begin{split}
\ln {\det}_{\rm ren}(1-eQ_iS\slashed V)  &= \frac{1}{2} \Pi_{VV} + \frac{1}{4}\Pi_{VVVV} \\
                                                                  &\quad + \ln {\det}_5 (1-eQ_iS\slashed V), 
\end{split}                                                                
\end{equation}
\begin{equation} \label{38}
\begin{split}
\ln {\det}_{5} &= {\rm Tr}\left[ \ln(1-eQ_iS\slashed V) + \sum_{n=1}^4 (eQ_iS\slashed V)^n \right].
\end{split}                                                                
\end{equation}

The sum of the vector fields' self-energy graphs $\Pi_{VV}$ formally corresponding to ${\rm Tr}(eQ_iS\slashed V)^2$ is calculated by expanding (\ref{34}) to $O(e^2,g^2,eg)$:

\begin{widetext}
\be \label{39}
\begin{split}
\Pi_{VV}  =  \frac{1}{4\pi^2} \int \frac{d^4k}{(2\pi)^4}~ \left| eQ_i \hat F_{\mu \nu}(k) + \frac{g_V^ig}{2\cos \theta_W} \hat Z_{\mu \nu}(k) \right|^2 \int_0^1 dz ~ z(1-z) \ln \left( \frac{z(1-z)k^2+m_i^2}{m_i^2} \right),
\end{split}
\ee
\end{widetext}

\noindent
where $\hat F$, $\hat Z$ denote Fourier transforms. By inspection of (\ref{39}) the transverse part of the photon self-energy $\Sigma^\gamma (k^2)$ and the photon-$Z$ mixing term $\Sigma^{\gamma Z}(k^2)$ from a charged fermion loop are normalized to vanish at $k^2=0$. The $ZZ$ term in (\ref{39}) is the transverse part of the neutral vector current contribution to the one-particle irreducible $Z$ self-energy $\Sigma^Z(k^2)$ from a charged fermion loop. The built-in renormalization subtractions in (\ref{34}) cause this contribution to vanish at $k^2=0$. When this contribution is combined with the remaining terms in $\Sigma^Z$ and continued to the Minkowski metric a finite mass renormalization counterterm $\delta M_Z^2$ can be chosen so that ${\rm Re}\Sigma^Z(k^2=M_Z^2) = 0$, where $M_Z$ is the pole mass.

The $\gamma \gamma$-scattering graph in (\ref{23}) is calculated from the vacuum polarization tensor $G_{\mu \nu \alpha \beta}$, where 

\begin{widetext}
\be \label{310}
\begin{split}
\Pi_{AAAA} = -(eQ_i)^4 \int d^4x_1 d^4x_2 d^4x_3 d^4x_4 ~ G_{\mu \nu \alpha \beta}(x_1,x_2,x_3,x_4) A^{\mu}(x_1) A^{\nu}(x_2) A^{\alpha}(x_3) A^{\beta}(x_4).
\end{split}
\ee
\end{widetext}

\noindent
$G_{\mu \nu \alpha \beta}$ is formally equal to $${\rm Tr}[ \gamma_\mu S(x_2-x_1)\gamma_\nu S(x_3-x_2)\gamma_\alpha S(x_4-x_3) \gamma_\beta S(x_1-x_4)]$$ and satisfies $\partial G_{\mu \nu \alpha \beta}/\partial x_{1\mu} =0$, etc. As noted above, reduction of this trace to $G_{\mu \nu \alpha \beta}$ is tedious but not necessary for our purpose here; all that is required is that this has somehow been done to give the unique result (\ref{310}). Then the shift $eQ_iA_{\mu} \to eQ_iA_{\mu} + g_V^igZ_{\mu}/(2\cos \theta_W)$ can be made in (\ref{310}) to give the expression for $\Pi_{VVVV}$ in (\ref{37}):

\be \label{311}
\begin{split}
\Pi_{VVVV} &= \Pi_{AAAA} + 4 \Pi_{AAAZ} + 4 \Pi_{AAZZ} \\
	           & \quad + 2\Pi_{AZAZ} + 4\Pi_{AZZZ} + \Pi_{ZZZZ}.
\end{split}
\ee

\noindent
The weight factors in (\ref{311}) indicate that the 16 terms in $\Pi_{VVVV}$ have been grouped together when possible using the symmetry properties of $G_{\mu \nu \alpha \beta}$.  It only remains to show that the potentially destabilizing growth of $\ln \det_{\rm ren}$ in (\ref{37}) as $|eQ_i|\mathcal{F} \to \infty$ due to $\Pi_{AAAA}$ and $\Pi_{AAAZ}$ does not occur. 

This is straightforward. Refer to the strong field growth of $\ln \det_{\rm ren}$ in (\ref{29}). All that is required is the shift $eQ_iF_{\mu \nu} \to eQ_i F_{\mu \nu} + g_V^igZ_{\mu \nu}/(2\cos \theta_W)$. In the absence of zero modes the right-hand side of (\ref{37}) behaves for large amplitude variations of $A_{\mu}$, and hence $F_{\mu \nu}$, for fixed $Z_{\mu}$ as

\be \label{312}
\begin{split}
{\ln \det}_{\rm ren} \qquad &\myeq \quad \frac{1}{48\pi^2} \norm{ eQ_i F_{\mu \nu} + \frac{g_V^ig}{2\cos \theta_W} Z_{\mu \nu}}^2 \\
				      & \qquad \times \ln \left( \frac{|eQ_i|\mathcal{F}}{m_i^2} \right) + R.
\end{split}
\ee

\noindent
The remainder $R$ continues to satisfy (\ref{210}). It is evident from (\ref{312}) that the $O(eQ_i\mathcal{F})^4$ and $O(eQ_i\mathcal{F})^3$ box graphs $\Pi_{AAAA}$ and $\Pi_{AAAZ}$ are cancelled by $\det_5$ in (\ref{37}) in this limit.

The asymptotic behavior seen in the $N$-independent term in (\ref{29}) for large $|eQ_i|\mathcal{F}$ was derived in Sections IV and VI of \cite{3}. The calculation of the asymptotic behavior in (\ref{312}) follows precisely the analysis in \cite{3} by replacing $eQ_iA_\mu$ in (\ref{25}) with the superposition in $eQ_iV_\mu$, resulting in (\ref{34}). The scaling parameter $|eQ_i|\mathcal{F}$ used in \cite{3} is now replaced with the scaling parameter $|eQ_i|\mathcal{F} + |g_{\,\, V}^i| g \mathcal{Z} / \cos \theta_W$, where $\mathcal{Z}$ is the amplitude of $Z_{\mu}$, giving it the dimension of $L^{-2}$. Letting this scaling parameter become large, whether due to the growth of $A_\mu$ or $Z_\mu$, results in a modified version of (\ref{312}) with the logarithm replaced with 
\begin{equation*}
    \ln \left( \frac{  |eQ_i|\mathcal{F} + |g_{\,\, V}^i|g\mathcal{Z} / \cos \theta_W}{m_i^2} \right).
\end{equation*}
For large amplitude variations of $A_\mu$ this reduces to (\ref{312}), with the remainder $R$ receiving a contribution of 
\begin{equation*}
    O\left( \frac{|eQ_i g_{\,\, V}^i| g ||F_{\mu \nu}||^2 \mathcal{Z}}{\mathcal{F}\cos \theta_W} \right)
\end{equation*}
so that $R$ continues to satisfy (\ref{210}).

\subsection{Zero modes} \label{SUB3B}

The purpose of this section is to state at least one of the cases for which $\ln \det_3$ in (\ref{34}) can assume a large negative value.

Following (\ref{211}) the operator $T_V$ in (\ref{36}) on which $\ln \det_3$ depends has non-vanishing eigenvalues $\{ \lambda_k \}_{k=1}^{\infty}$ obtained from $T_V | \lambda_k \ket = \lambda_k | \lambda_k \ket$ by transforming this into the equivalent equation

\be \label{313}
\begin{split}
\left[ (P - eQ_iV)^2 + \frac{eQ_i}{2\lambda_k}\sigma_{\mu \nu} V^{\mu \nu} \right] \psi_{ {\lambda_k} ,n} = -m_i^2  \psi_{ {\lambda_k} ,n},
\end{split}
\ee

\noindent
where $V_{\mu}$ is given by (\ref{31}) and $ \psi_{ {\lambda_k} ,n} \in L^2(\mathbb{R}^4)$. We continue to use the same notation as in Section \ref{SUB2C}. Equation (\ref{313}) is simply the quantum mechanical problem of finding the values of $\lambda_k$ that result in a bound state of the Hamiltonian on the left-hand side with energy $-m_i^2$. It makes no reference to $V_\mu$ being a superposition of $A_{\mu}$ and $Z_{\mu}$. It is assumed that $eQ_i >0$ and that $0<\lambda_k<1$ which requires that $\bra \lambda_k, n | \sigma_{\mu \nu} V^{\mu \nu} | \lambda_k,n \ket< 0$. Recall that the eigenvalues occur in pairs that satisfy the bound $|\lambda_k|<1$. The analysis in \cite{3} leading to this result extends to the electroweak model since it only requires that $V_{\mu}$ is a neutral vector field. The state $| \lambda_k,n \ket$ will generally have mixed chirality. Proceeding as in Section \ref{SUB2C} suppose that the potential $A_{\mu}$ in (\ref{31}) also supports a zero mode $|0,l\ket$ with definite chirality that satisfies (\ref{212}) and hence has $\bra 0,l | \sigma_{\mu \nu} F^{\mu \nu} | 0, l \ket < 0$. The state $|0,l \ket$ denotes a zero mode state with quantum numbers $l$ and not a state with $\lambda_k = 0$. Then from (\ref{313}), 

\be \label{314}
\begin{split}
\bra 0,l |(P-eQ_iV)^2 + \frac{eQ_i}{2\lambda_k} \sigma_{\mu \nu} V^{\mu \nu} | \lambda_k,n \ket = -m_i^2 \bra 0,l | \lambda_k, n\ket.
\end{split} 
\ee

\noindent
We have remarked that the zero mode states $\psi_{0,n}(x)$ have a distinctive structure, and so we expect that $\bra 0, l | \lambda_k,n \ket \neq 0$ only for a particular $\lambda_k$, say $\lambda$, and only if the states' quantum numbers $l=n$. Suppose this to be the case.

Form the inner product (\ref{212}) with $| \lambda, n \ket$ and subtract the complex conjugate of (\ref{314}) from it to obtain

\begin{widetext}
\be \label{315}
\begin{split}
\left( \frac{1-\lambda}{\lambda}\right) |eQ_i| \left| \frac{ \bra \lambda, n | \sigma_{\mu \nu} F^{\mu \nu} | 0,n \ket }{\bra \lambda, n  | 0,n \ket} \right|  \leq 4|c|   \left| \frac{ \bra \lambda, n | Z(P-eQ_iA)| 0,n \ket }{\bra \lambda, n  | 0,n \ket} \right| + 2  \left| \frac{ \bra \lambda, n | ic\partial_{\mu} Z^{\mu} + c^2Z^2 + \tfrac{c}{2\lambda} \sigma_{\mu \nu}Z^{\mu \nu} + m_i^2 | 0,n \ket }{\bra \lambda, n  | 0,n \ket} \right|,
\end{split}
\ee
\end{widetext}

where $c= g_V^ig/(2\cos \theta_W)$. The upper bound in (\ref{315}) is gauge invariant in $A_{\mu}$ by inspection. Note that 

\begin{widetext}
\be \label{316} 
\begin{split}
| \bra \lambda, n | Z(P-eQ_iA) | 0,n \ket |  & \leq ( \bra \lambda, n | Z^2 | \lambda,n \ket )^{1/2} | \bra 0,n | (P-eQ_iA)^2 | 0,n \ket |^{1/2} \\
							       & \leq  ( \bra \lambda, n | Z^2 | \lambda,n \ket )^{1/2}  | \bra 0,n | -\tfrac{1}{2} eQ_i \sigma_{\mu \nu} F^{\mu \nu} | 0,n \ket |^{1/2},
\end{split}
\ee
\end{widetext}

\noindent
where we used the Schwarz inequality and (\ref{212}). Then (\ref{315}) can be rewritten as 

\be \label{317}
\begin{split}
\left( \frac{1-\lambda}{\lambda} \right) |eQ_i| \mathcal{F} K_1 \leq (|eQ_i|\mathcal{F})^{1/2} K_2 + K_3,
\end{split} 
\ee
where we define 

\begin{equation} \label{318}
K_1 =  \left| \frac{ \bra \lambda, n | \sigma_{\mu \nu} F^{\mu \nu} / \mathcal{F} |0, n  \ket }{\bra \lambda, n | 0, n \ket} \right| ,
\end{equation}
\begin{equation} \label{319}
K_2 = \frac{4|c| (\bra \lambda, n | Z^2 | \lambda, n \ket)^{1/2} | \bra 0,n | \sigma_{\mu \nu} F^{\mu \nu}/(2\mathcal{F}) | 0, n \ket |^{1/2} }{ | \bra \lambda, n | 0, n \ket | }
\end{equation}
\vspace{0.1em}

\noindent
and $K_3$ is the second term on the right-hand side of (\ref{315}). The constants $K_1, K_2, K_3$ are bounded for large $\mathcal{F}$ and have dimension $L^0, L^{-1}, L^{-2}$, respectively.

To solve for $\lambda$ let $\lambda = 1-\delta$ to obtain 

\begin{equation} \label{320}
\begin{split}
\delta &\leq \frac{ K_2/K_1 }{ (|eQ_i|\mathcal{F})^{1/2} } + \frac{K_3/K_1 - (K_2/K_1)^2}{|eQ_i|\mathcal{F}}\\
&\quad + O(1/(|eQ_i|\mathcal{F} )^{3/2}),
\end{split} 
\end{equation}

\noindent
and

\begin{equation} \label{321}
\ln (1-\lambda^2) \leq - \ln  \left[ \left( \frac{K_1}{2K_2} \times (|eQ_i|\mathcal{F})^{1/2} \right) \right] + O(1/(|eQ_i|\mathcal{F} )^{1/2}).
\end{equation}

\noindent
From (\ref{27}) on substituting $T$ with $T_V$,

\begin{widetext}

\begin{equation} \label{322}
\begin{split}
\ln {\det}_3 (1+T_V) & = \sum_{k=1}^{\infty} [\ln (1-\lambda_k^2) + \lambda_k^2] \\
			       & \leq -\frac{N}{2} \left[ \ln\left( \frac{ |eQ_i|\mathcal{F} }{M_i^2} \right) - 2 + O(1/(|eQ_i|\mathcal{F})^{1/2} \right]+ \sum_{\lambda_k \neq \lambda} [\ln (1-\lambda_k^2) + \lambda_k^2],
\end{split} 
\end{equation}
\end{widetext}

\noindent
where we define $M_i = 2K_2/K_1$, giving $M_i$ dimension $L^{-1}$. $N$ is the degeneracy of the zero-mode states $|0,n\ket$. Following Section \ref{SUB2C},  $N = (eQ_i)^2 | \int d^4x ~ \epsilon_{\mu \nu \alpha \beta} F^{\alpha \beta} F^{\mu \nu} | / (32\pi^2)$ when all of the zero mode states have the same chirality. Referring to (\ref{34}), the result (\ref{322}) multiplied by $1/2$ modifies the result (\ref{312}) to 

\begin{widetext}
\begin{equation} \label{323}
\begin{split}
{\ln \det}_{\rm ren} \qquad & \myleqq \qquad \frac{(eQ_i)^2}{48\pi^2} \norm{ F_{\mu \nu} }^2 \ln \left(  \frac{ |eQ_i|\mathcal{F} }{m_i^2} \right) - \frac{N}{4} \ln \left( \frac{|eQ_i|\mathcal{F}}{M_i^2} \right) + \frac{1}{2} \sum_{\lambda_k \neq \lambda} [\ln (1-\lambda_k^2) + \lambda_k^2] + R.
\end{split} 
\end{equation}
\end{widetext}

$R$ continues to satisfy (\ref{210}) when all of the zero mode states have the same chirality; otherwise we cannot place a bound on $R$. 

The sum of the remaining eigenvalues in (\ref{323}) is convergent and negative. It is possibly the most critical contribution to $\ln \det_{\rm ren}$. Even if a potential does not support a zero mode - thereby removing the $N$-dependent term in (\ref{323}) - we know of no a priori reason why the eigenvalues satisfying $| \lambda_k(\mathcal{F} \to \infty) | <1$ should sum to a bounded function of $\mathcal{F}$. It should be kept in mind that the eigenvalue problem in (\ref{313}) is equivalent to finding the bound states of a charged fermion in a random four-dimensional magnetic field. There is apparently no limit to the complexity of magnetic fields generated by the Maxwell measure $d\mu(A)$ for fixed $Z_{\mu}$. Although the degeneracy associated with each of the eigenvalues $\{ \lambda_k \}_{k=1}^{\infty}$ may be a slowly varying function of $\mathcal{F}$, their sum may compete with the leading term in (\ref{323}) whose sign is determined by QED's lack of asymptotic freedom.

\subsection{Remaining determinants}\label{SUB3C}

We return to the second determinant in (\ref{32}) depending on the axial vector current and the Higgs field. Making mathematical sense of this determinant is a large problem that will have to be dealt with in a subsequent paper. In order to renormalize it 2 tadpole, 7 two-point, 16 triangle and 31 box graphs have to be factored out. Of the 16 triangle graphs four are anomaly bearing and cancel when summed over generations of fermions. Seven of the triangle graphs vanish by Euclidean $C$-invariance. The 5 remaining graphs are Higgs field dependent. The 6 anomaly-bearing box graphs also vanish by $C$-invariance, including the potentially destabilizing graph $AAAZ\gamma_5$; the non-anomalous Higgs graph $AAAH$ likewise vanishes by $C$-invariance. These calculations have been completed with the assumption that $H$ has also been smoothed as in Section \ref{SUB2A} with an ultraviolet cutoff parameter different from $\Lambda$ and $\tilde \Lambda$ used for $A$ and $Z$, respectively.

It remains to place a bound on the absolutely convergent remainder of the second determinant as $|eQ_i|\mathcal{F}\to \infty$. The leading term is the pentagon graph

\begin{equation} \label{324}
\begin{split}
\Pi_5 = (eQ_i)^4 {\rm Tr} \left[ S \slashed VS \slashed VS \slashed VS \slashed V S_V \left( \frac{gg_A^i}{2\cos \theta_W} \slashed Z \gamma_5 + \frac{gm_i}{2M_W}H \right) \right],
\end{split} 
\end{equation}

\noindent
where $S_V$ is from (\ref{33}). This graph is absolutely convergent since 
$S_V$'s short-distance behavior is less singular than in the free field case for the class of potentials considered here. This conclusion is reached by approximating the local field lines by a constant field and noting the enhanced propagation of $S_V(x',x'')$ parallel to the field lines, resulting in a short-distance behavior of $1/(x'-x'')_{||}^2$.
It was found in \cite{1} that when $S_A = (\slashed p - e Q_i \slashed A + m_i)^{-1}$ occurs in an absolutely convergent fermion loop $S_A$'s effective falloff for large variations of $A_{\mu}$ induced by the scaling $A_{\mu} \to LA_{\mu}$ is $O(1/L^2)$ when $L\to \infty$ for $A_{\mu} \in \cap_{r>4} L^r(\mathbb{R}^4)$. The analysis leading to this result relied on (\ref{23}) and (\ref{29}) with $R$ satisfying (\ref{210}). The one-to one correspondence between (\ref{23}), (\ref{37}) and (\ref{29}), (\ref{312}) with $R$ also satisfying (\ref{210}) allows the same conclusion to be drawn about $S_V$. A large variation of $V_\mu$ can be induced by a large variation of $A_\mu$ so that $V_\mu \to L( A_{\mu} + g_V^i Z_\mu / (2LQ_i \cos \theta_W \sin \theta_W)^{-1}) \equiv L \tilde V_{\mu}$. Then $S_V \to S_{L\tilde V} = (\slashed P - eQ_i L \slashed{\tilde V} + m_i)^{-1}$ and 

\begin{equation} \label{325}
\begin{split}
\Pi_5 & \to (eQ_iL)^4  {\rm Tr} \left[ S \slashed {\tilde V}S \slashed  {\tilde V}S \slashed  {\tilde V}S \slashed  {\tilde V} S_{L\tilde V} \left( \frac{gg_V^i}{2\cos \theta_W} \slashed Z \gamma_5 + \frac{gm_i}{2M_W}H \right) \right] \\
&= O(L^2) \
\end{split} 
\end{equation}

\noindent
for fixed $Z_{\mu}$ and $H$ following the above result for $S_A$. Then $\Pi_{5}$ grows at most quadratically for large variations of $A_{\mu}$. There are still some technical difficulties that have to be resolved in order to bound all of the second determinant's remainder.

The last determinants to consider are those contributed by the  hadronic and leptonic sectors of the charged weak-vector current. The hadronic determinant contributed by quark $i$, without mixing, corresponds to the $W$-dependent determinant in (\ref{12e}).  The leptonic determinant is obtained from this by the exchanges $W^+ \leftrightarrow W^-$ and $t_{3L}(i) \to -t_{3L}(i)$.  To illustrate the effect of mixing it suffices to consider the two generations $u,d$ and $c,s$. Let it be decided to integrate over the $u$ and $c$ quarks first followed by integration over $d$ and $s$ quarks. The result is the product of determinants

\begin{widetext}
\begin{equation} \label{326}
\begin{split}
&\det \Bigg[ 1 - \frac{g^2}{8} G_d \slashed W^{-} (1-\gamma_5) (|V_{ud}|^2G_u + |V_{cd}|^2G_c )\slashed W^{+} (1-\gamma_5)  \Bigg]  \\
\times & \det \Bigg[ 1 - \frac{g^2}{8} G_s \slashed W^{-} (1-\gamma_5) (|V_{us}|^2G_u + |V_{cs}|^2G_c )\slashed  W^{+} (1-\gamma_5) \\
 & \qquad - \frac{g^4}{64} G_sO_uG_d ( O_u |V_{us}|^2 |V_{ud}|^2 + O_c V_{us}^* V_{ud} V_{cd}^*V_{cs}) \\
 & \qquad - \frac{g^4}{64} G_sO_cG_d ( O_c |V_{cs}|^2 |V_{cd}|^2 + O_u V_{cs}^* V_{cd} V_{ud}^*V_{us}) \Bigg], 
\end{split} 
\end{equation}
\end{widetext}

\noindent
where

\begin{widetext}
\begin{eqnarray} \label{327}
O_i &=& \slashed W^- (1-\gamma_5) G_i \slashed W^+ (1-\gamma_5) \\ \label{328}
G_i &=& \left( \slashed P + m_i - eQ_i\slashed A - \frac{g}{2\cos \theta_W} \slashed Z (g_V^i - g_A^i\gamma_5) + \frac{gm_i}{2M_W} H \right)^{-1},
\end{eqnarray}
\end{widetext}

\noindent
$g_V^i, g_A^i$ are given by (\ref{14}), (\ref{15}) and $V_{ij}$ are the Cabibbo-Kobayashi-Maskawa matrix elements \cite{2}. Including the $t$ and $b$ quarks results in strings of up to eight propagators. 

As we are interested in the growth of these determinants for large amplitude variations of $A_{\mu}$ it is advantageous to factor out $A_{\mu}$ from the propagator $G_i$ through the operator identities

\begin{widetext}
\begin{eqnarray} \label{329}
G_i &=& S_A - S_A \left[ -\frac{g}{2\cos \theta_W} (g_V^i - g_A^i \gamma_5) \slashed Z + \frac{gm_i}{2M_W}H \right]G_i \\ \label{330}
S_A&=& S + SeQ_i \slashed A S_A. 
\end{eqnarray}
\end{widetext}

\noindent
Comparing (\ref{326}) with the unmixed quark determinant in (\ref{12e}) it is seen that mixing only adds more propagators which tend to suppress the growth of the added terms. This follows from (\ref{329}) and (\ref{330}) and the falloff of $S_A$ for large amplitude variations of $A$. So we believe it is safe to neglect quark and neutrino mixing for the purpose of this paper. Instead we focus on the unmixed hadronic determinant

\begin{widetext}
\begin{equation}\label{new51}
{\ln \det}\left[1-\frac{g^2}{8}  G_{t_{3L}(i)=- \frac{1}{2}}^{} {\slashed W}^{-}(1-\gamma_5)G_{t_{3L}(i)= \frac{1}{2}}  {\slashed W}^{+}(1-\gamma_5)\right],
\end{equation}
\end{widetext}
and its leptonic sister determinant.

Terms of $O(g^2, g^3, g^4, eg^2, eg^3, g^2e^2)$ have to be factored out of (\ref{new51}). This is done by iterating (\ref{330}) twice and substituting the result in (\ref{329}) which in turn is inserted in (\ref{new51}) and the leptonic determinant, followed by a loop expansion. These terms have to be renormalized and the chiral anomalies cancelled by summing over generations. The example of the triangle graph $W^+W^-\gamma$ is given in \cite{3}. It is assumed that the $W^{\pm}$ fields have been smoothed following \ref{SUB2A} with an ultraviolet cutoff different from that of $A, Z$ and $H$. The remaining loop graphs are absolutely convergent, and all contain $S_A$. The leading remaining graphs have the general form 
\begin{equation*}
{\rm Tr}[S \slashed AS \slashed A S \slashed A S \slashed W^- (1-\gamma_5)S_A \slashed W^+(1-\gamma_5)]
\end{equation*} 
\noindent
plus permutations of $S$ and $S_A$. If $A$ is scaled by $L$ the effective falloff of $S_A$ is $O(1/L^2)$ for $L \to \infty$ \cite{1}, and so these graphs are $O(L)$. Of course, the entire remainder has to be shown to be $O(L^2)$ or less for the Maxwell integration in (\ref{11}) to have a chance of converging.

The above remarks on the falloff of $A_\mu$ also apply to $Z_\mu$ as this is essential to the analysis of Sections \ref{SUB3A}-\ref{SUB3C}.

\subsection{Volume cutoff}\label{SUB3D}

The constant $\mathcal{N}$ in (\ref{11}) is obtained by setting $\mathcal{I} = \mathcal{F} = 1$ so that 

\begin{widetext}
\begin{equation} \label{331}
\frac{1}{\mathcal{N}} \int d\mu(A) d\mu(Z) d\mu(W^\pm) d\mu(H) ~ \exp\left[ \sum_i \ln \det(i) + \int d^4x ~ \mathcal{L}_{\rm int}(A,Z,W^\pm,H) \right] = 1.
\end{equation}
\end{widetext}

\noindent
The first term in the effective action in the exponential is the sum of the two determinants on the right-hand side of (\ref{32}), 
the hadronic determinant in (\ref{new51}) and its leptonic counterpart.
These are assumed to be renormalized and freed of anomalies as outlined in Sections \ref{SUB3A}-\ref{SUB3C} above. The second term, $\mathcal{L}_{\rm int}$, consists of the gauge boson and Higgs self-couplings \cite{21}. Then $\mathcal{N}$ is seen to be a normalization constant that makes $d\mu(A) d\mu(Z) \ldots \exp[\cdot]$ a probability measure. The integral in (\ref{331}) generates the vacuum self-energy and should cancel in the calculation of a physical process $\mathcal{I}$ in (\ref{11}). 

Even if all of the functional integrals in (\ref{331}) converge the translation invariance of the vacuum self-energy introduces a volume divergence causing $\mathcal{N} = \infty$. Unless this divergence is controlled one is dealing with senseless functional integrals that are not subject to mathematical analysis. The scope of this paper only requires that we deal with the Maxwell integration. The effective action is gauge invariant and is dependent on $F_{\mu \nu}$ only. A gauge invariance preserving volume cutoff can be introduced in principle by replacing $F_{\mu \nu}$ everywhere with $gF_{\mu \nu}$, where $g$ is a space cutoff such as $g\in C_0^\infty$. Implementing this requires knowledge of the explicitly gauge invariant form of the Maxwell sector of the one-loop effective action. At present this is limited to its strong field limit and its large fermion mass expansion.

For the results reported here to be relevant a typical distributional, random connection $A_{\mu}$ should have $\mu(A)$-measure 1. It has been assumed that $A_\mu$ falls off at least as fast as $1/|x|$ as $|x| \to \infty$ following [36]. To the author's knowledge there are no results for a typical $A$'s large-distance behavior. It is known that a typical $\varphi(x)$ supporting the four-dimensional Gaussian measure $d\mu(\varphi)$ of a free, massive, spin-$0$ boson has growth $|x|^2 (\ln|x|)^\beta$, $\beta >1/2$ \cite{22,23}. If a typical $A_\mu$ does not fall off as assumed here the presence of a smoothly decaying volume cutoff such as $g$ would allow the continued applicability of the theorems used to obtain the above results. Accordingly, they will remain intact when the volume cutoff is fully implemented. 

The above remarks on the falloff of $A_{\mu}$ and those in footnote [36] also apply to $Z_{\mu}$ as this is essential to the analysis of Sections \ref{SUB3A}-\ref{SUB3C}.

\section{Conclusion} \label{SECTION4}

It has been shown that the box graphs $AAAA$ and $AAAZ$ do not obstruct the non-perturbative path integral quantization of the electroweak model. This is subject to the provision that the fermion degrees of freedom are first integrated out to obtain an effective action followed by the functional integral over the Maxwell field. These box graphs are present in the effective action and cannot be avoided. There are other potentially destabilizing terms in the effective action contributed by the axial vector current, the charged vector current and the Higgs field. Based on a previous result for the falloff of $S_A^{}  = (\slashed P-eQ_i \slashed A + m_i)^{-1}$   when the random potential's amplitude is large \cite{1} we expect these terms to be subleading compared to the strong-field behavior of the QED effective action. In this sense we can say QED decouples from the rest of the electroweak model in this limit.

Evidence has been given that zero mode bearing potentials supporting the functional measure $d\mu (A)$, if any, are highly relevant to the convergence of the Maxwell field integration in (\ref{11}). The weight assigned to these potentials by $d\mu (A)$ has been an open question for at least forty years. And so it remains.

The paramagnetic term - $\ln \det_3$ - in the Maxwell sector of the one-loop effective action may be critical to the convergence of the functional integral over $A$, whether or not zero bearing potentials are supported by $d\mu(A)$. Deciding the issue depends on finding the degeneracy factors in $\ln \det_3$'s eigenvalue expansion in (\ref{27}) and (\ref{323}) and their dependence on the large amplitude limit of the random magnetic fields generated by $d\mu(A)$ for fixed $Z_\mu$.

\appendix

\section{}
The estimate of the growth of the scalar QED renormalized determinant, $\ln \det_{SQED}$, for large amplitude variations of $F$ was completed in \cite{3} up to a remainder $R$ defined below. Here we wish to verify that $R$ is subdominant. From (3.3) in \cite{3} 
\begin{widetext}
\begin{eqnarray}\label{A1}
\ln {\rm det}_{\rm SQED} &=& \int_0^\infty \frac{dt}{t} \left[ {\rm Tr} \left( e^{-P^2t} - e^{-(P-eA)^2t} \right)  - \frac{e^2||F^2||}{192\pi^2}\right] e^{-tm^2} \\ \nonumber
                         &=& \int_0^{1/e\mathcal{F}}  \frac{dt}{t} \left[ {\rm Tr} \left( e^{-P^2t} - e^{-(P-eA)^2t} \right)  - \frac{e^2||F^2||}{192\pi^2}\right]e^{-tm^2} \\ \label{A2}
                         && - \frac{e^2 ||F^2||}{192\pi^2} \int_{1/e\mathcal{F}}^\infty \frac{dt}{t} e^{-tm^2}+R
\end{eqnarray}
where
\begin{eqnarray} \label{A3}
R &=& \int_{1/e\mathcal{F}}^{\infty} \frac{dt}{t} {\rm Tr} \left( e^{-P^2t} - e^{-(P-eA)^2t} \right) e^{-tm^2} \\ \label{A4}
  & \geq& 0,
\end{eqnarray}
\end{widetext}
following Kato's inquality in the form \cite{25,26,27}
\begin{equation}\label{A5}
{\rm Tr}\left( e^{-P^2t} - e^{-(P-eA)^2t} \right) \geq 0.
\end{equation}
Without explicitly calculating $R$ we obtained 
\begin{equation}\label{A6}
\ln {\rm det}_{\rm SQED} \geq - \frac{e^2||F^2||}{192\pi^2} \ln\left( e\mathcal{F} / m^2 \right) + O(e\mathcal{F})^2, 
\end{equation}
where the inequality sign follows from (\ref{A2}) and (\ref{A4}). We will now estimate $R$ in order to turn (\ref{A6}) into an equality. 
\begin{widetext}
Consider
\begin{eqnarray} \nonumber
    \partial R / \partial e &=& e^{-1} {\rm Tr}\left( e^{-P^2/e\mathcal{F}} - e^{-(P-eA)^2/e\mathcal{F}} \right) e^{-m^2/e\mathcal{F}} \\ \label{A7}
                             && - \int_{1/e\mathcal{F}}^\infty \frac{dt}{t} \frac{\partial}{\partial e} {\rm Tr}\left( e^{-(P-eA)^2t} \right)e^{-tm^2},
\end{eqnarray}
with $R(e=0)=0$. The first term in (\ref{A7}) can be estimated by making a heat kernel expansion. Define 
\begin{eqnarray}\label{A8}
    e \frac{\partial R_1}{\partial e} &=&   {\rm Tr}\left( e^{-P^2/e\mathcal{F}} - e^{-(P-eA)^2/e\mathcal{F}} \right) e^{-m^2/e\mathcal{F}} \\ \label{A9}
    & \myqeq & \qquad \sum_{n=0}^N a_n(eF)(1/e\mathcal{F})^n + a_M(eF)(1/e\mathcal{F})^M,
\end{eqnarray}
\end{widetext}
where $a_M$ is the first nonzero coefficient after $a_N$, assuming that the expansion is an asymptotic series \cite{30}. The first few terms of the series are \cite{3}
\begin{widetext}
\begin{equation}\label{A10}
\begin{split}
   & \frac{1}{16\pi^2} \int d^4x \Big\{ \frac{e^2}{12} F_{\mu \nu}F^{\mu \nu} + \frac{e^2}{120} F_{\mu \nu} \nabla^2 F^{\mu \nu}/(e\mathcal{F}) + \frac{e^2}{1680} F_{\mu \nu} \nabla^4 F^{\mu \nu}/ (e\mathcal{F})^2 \\
   & + \frac{e^4}{1440} \left[ (\tilde F_{\mu \nu} F^{\mu \nu})^2 - 7(F_{\mu \nu}F^{\mu \nu})^2\right]/(e\mathcal{F})^2 \Big\} + O(1/(e\mathcal{F})^3),
\end{split}
\end{equation}
\end{widetext}
where $\tilde F_{\mu \nu}= \epsilon_{\mu \nu \alpha \beta} F^{\alpha \beta}/2$.

The expansion in (\ref{A9}) assumes that $F$ is infinitely differentiable, which it is since it is calculated from the smooth potentials $A$ introduced in Section \ref{SUB2A}. The validity of (\ref{A9}) also requires that all of the trace terms over $F_{\mu \nu}$ and its derivatives converge, which they do if we assume $F_{\mu \nu}\in \cap_{r=2} L^r(\mathbb{R}^4)$. We have previously assumed $r>2$; this will be discussed at the end of this Appendix.

Since $[1/(e\mathcal{F})] = L^2$, the maximum power of $F$ in $a_M$ is $M+2$ so that the truncation error in (\ref{A9}) never exceeds $(e\mathcal{F})^2$.

Rewrite the expansion coefficients in terms of the dimensionless scaled feld $f_{\mu \nu}$ defined by $F_{\mu \nu}(x) =\mathcal{F} f_{\mu \nu}(x)$. The scale factor $\mathcal{F} \equiv {\rm max}_x \sqrt{F_{\mu \nu}(x) F^{\mu \nu}(x) } = {\rm max}_x |F_{\mu \nu}(x)|$. Introduce the amplitude $L$ of $A_{\mu}$ by $L = {\rm max}_x|A_\mu(x)|$ so that $A_\mu(x) = L a_{\mu}(x)$, where $a_\mu$ is dimensionless and $|a_\mu(x)| \leq 1$. Then $\mathcal{F} = L {\rm max}_x |\partial_\mu a_\nu(x) - \partial_\nu a_\mu(x)|$ and hence $|f_{\mu \nu}(x)|\leq 1$. Terms in the series containing gradients $\nabla^{2n}$ will scale as $\nabla^{2n}/(e\mathcal{F})^n$ when factoring out $e\mathcal{F}$, and therefore will be subdominant as $e \mathcal \to \infty$.  Hence, the series in (\ref{A9}) has the form 
\begin{widetext}
\begin{equation}\label{A11}
e \frac{\partial R_1}{\partial e}\quad \myqeq \quad (e\mathcal{F})^2 \left( \sum_{n=0}^N a_n(f,\nabla^{2m}f/(e\mathcal{F})^m) + a_M(f, \nabla^{2m}f/(e \mathcal{F})^m) \right).
\end{equation}
\end{widetext}
The only certain statement we can make about this series is that for any $N < \infty$ 
\begin{equation}\label{A12}
    R_1 \quad \myqeq \quad O(e\mathcal{F})^2.
\end{equation}

Consider the second term in (\ref{A7}). Define 
\begin{equation}\label{A13}
   \frac{\partial R_2}{\partial e} = - \int_{1/e\mathcal{F}}^{\infty} \frac{dt}{t}\frac{\partial}{\partial e} {\rm Tr} \left( e^{-(P-eA)^2t} \right) e^{-tm^2}.
\end{equation}
The Wiener path integral representation of the integral kernel is \cite{extra}
\begin{equation}\label{A14}
    e^{-(P-eA)^2t(x,y)} = \int_{\omega} \exp \left( -ie \int_{\omega(0)=x}^{\omega(t)=y} A(\omega)d\omega  \right)d\mu_{x,y,t}(\omega),
\end{equation}
where the free Wiener measure in four dimensions gives the probability density of finding a particle at $y$ at time $t$ if it started at $x$ at $t=0$,
\begin{eqnarray}\nonumber
    \int_{\omega} d\mu_{x,y,t} (\omega) &=& e^{-tP^2}(x,y) \\ \label{A15}
                                        &=& (4\pi t)^{-2}e^{-(x-y)^2/4t}.
\end{eqnarray}
Then
\begin{widetext}
\begin{equation}\label{A16}
    \frac{\partial}{\partial e} {\rm Tr }\left( e^{-(P-eA)^2t} \right) = -i \int d^4x\, \int_\omega \oint A(\omega) d\omega \, e^{-ie \oint A(\omega)d\omega} d\mu_{x,x,t}(\omega).
\end{equation}
\end{widetext}
We take it as a reasonable assumption that the right-hand side of (\ref{A16}) vanishes as $e \to \infty$ due to the rapidly oscillating exponential of $A$, and hence
\begin{equation}\label{A17}
     \frac{\partial}{\partial e} {\rm Tr }\left( e^{-(P-eA)^2t} \right)\quad \myqeq \quad 0.
\end{equation}
Since $R_2$ is gauge invariant $e$ always appears in the combination $e F_{\mu \nu}$.

Temporarily rescaling $R$ in (\ref{A3}) by letting $tm^2 = u$ we see that the lower limit transforms to $m^2/e\mathcal{F}$, indicating that letting $m \to 0$ is consistent with $e\mathcal{F} \to \infty$.
As there are no zero modes in scalar QED and the on-shell charge renormalization of ${\rm det}_{\rm SQED}$ has already been made in (\ref{A6}), the $m=0$ limit can be taken in (\ref{A13}). This requires that the right-hand side of (\ref{A17}) behaves as 

\begin{equation}\label{A18}
    \frac{\partial}{\partial e} {\rm Tr }\left( e^{-(P-eA)^2t} \right)\quad \myqeq \quad  e^{-c(e\mathcal{F}t)^\propto} \frac{\partial}{\partial e}   \int d^4x ~ g(eF,t),
\end{equation}
for some $c,\propto >0$. The integrand, $g$, must result in a convergent integral and have dimension $L^{-4}$. These requirements restrict $g$ to be a linear function of the variables $[(eF)^n]_{\mu \nu} \nabla^{2l} [(eF)^m]^{\mu \nu} t^p$ with 
\begin{equation}\label{A19}
    n+m+l-p = 2; \quad l \geq 1 \quad {\text{ when }} m \geq 1
\end{equation}
for $l,m,n \in \mathbb{Z}^+$, $p\geq 0$. On replacing $F$ with $\mathcal{F}f$, $g$ has the scaling property
\begin{equation}\label{A20}
\begin{split}
   & g([(eF)^n]_{\mu \nu} \nabla^{2l} [(eF)^m]^{\mu \nu} t^p) =  \\
   &  \qquad (e\mathcal{F})^2 g((f^n)_{\mu \nu} \nabla^{2l} (f^m)^{\mu \nu} (e\mathcal{F}t)^p/(e\mathcal{F})^p ).
\end{split}
\end{equation}
Substituting (\ref{A18}), (\ref{A20}) in (\ref{A13}) gives
\begin{widetext}
\begin{equation}\label{A21}
    \frac{\partial R_2}{\partial e} \qquad \myqeq \qquad -\int_{1/e\mathcal{F}}^\infty  \frac{dt}{t} e^{-c(e\mathcal{F}t)^{\propto}} \frac{\partial}{\partial e} \left[ (e\mathcal{F})^2 \int d^4x \, g( f^n\nabla^{2l}f^m (e\mathcal{F}t)^p / (e\mathcal{F})^p )  \right].
\end{equation}
\end{widetext}
The differentiation can be done by inspection. Substituting $z = e\mathcal{F}t$ and integrating gives
\begin{equation}\label{A22}
    R_2 \quad \myqeq \quad O(e\mathcal{F})^2,
\end{equation}
or less if the gradient terms in (\ref{A21}) dominate $g$. Combining (\ref{A7}), (\ref{A8}), (\ref{A13}) and the results (\ref{A12}) and (\ref{A22}) gives
\begin{equation}\label{A23}
    R \quad \myqeq \quad  O(e\mathcal{F})^2,
\end{equation}
with $R>0$. This result is almost certainly correct. For if it increased as $O(e\mathcal{F})^\beta$, $\beta >2$, then scalar QED's effective action would decrease as $S_{\rm SQED} = -\ln {\rm det}_{\rm SQED} \quad \myqeq \quad -O(e\mathcal{F})^{\beta}$. We have no evidence for this. There are no $(e\mathcal{F})^2 [\ln (e\mathcal{F}/m^2)]^k$, $k>0$, terms in $R$ for the reasons stated under (\ref{A17}).

When (\ref{A23}) is combined with (\ref{A2}) and (\ref{A6}), where $R$ is neglected, we can rewrite (\ref{A6}) as
\begin{equation}\label{A24}
    \ln {\rm det}_{\rm SQED} \quad \myqeq \quad -\frac{e^2||F^2||}{192\pi^2} \ln (e\mathcal{F}/m^2) + O(e\mathcal{F})^2.
\end{equation}
This result assumes that the distributional, random connection $A$ has been smoothed as in \ref{SUB2A} and that these smoothed potentials $A \in \cap_{r>4} L^r(\mathbb{R}^4)$. It has also been assumed throughout this paper that $F\in \cap_{r>2} L^r(\mathbb{R}^4)$. The term $||F||^2$ in (\ref{A24}) is from charge renormalization and is divergent under this assumption. It must be replaced with $||gF||$, where $g$ is the volume cutoff introduced in \ref{SUB3D}. Since $\det_5$ is related to ${\rm det}_{\rm SQED}$ through (\ref{23}) and (\ref{25}) and is finite without a volume cutoff we conjecture that any residual terms containing $||F||$ in the remainder $R$ cancel as they do in the remainder in (4.6) in \cite{3}. In this case the assumption $r>2$ is possible.



\end{document}